\pdfoutput=1

\documentclass[useAMS,usenatbib]{mn2e}

\usepackage{graphicx}
\usepackage{dcolumn}
\usepackage{bm}
\usepackage{amsmath}
\usepackage{amsfonts}
\usepackage{psfrag}
\usepackage{here}
\usepackage{rotating}
\usepackage{fixltx2e}
\usepackage{float}

\title[1.4 GHz on the Fundamental Plane]{1.4 GHz on the Fundamental Plane of Black Hole Activity}
\author[P. Saikia et al. 2018]{Payaswini Saikia \thanks{E-mail: p.saikia@astro.ru.nl}, Elmar K\"{o}rding \thanks{E-mail: e.koerding@astro.ru.nl} and Salome Dibi\\
Department of Astrophysics/IMAPP, Radboud University, Nijmegen, P.O. Box 9010, 6500 GL Nijmegen, The Netherlands}

\begin{document}

\date{Accepted 2018 March 19. Received 2018 March 19; in original form 2017 July 11}

\pagerange{\pageref{firstpage}--\pageref{lastpage}} \pubyear{2018}

\maketitle

\label{firstpage}


\begin{abstract}
The fundamental plane of black hole activity is an empirical relationship between the OIII/X-ray luminosity depicting the accretion power, the radio luminosity as a probe of the instantaneous jet power and the mass of the black hole. For the first time, we use the 1.4 GHz FIRST radio luminosities on the optical fundamental plane, to investigate whether or not FIRST fluxes can trace nuclear activity. We use a SDSS-FIRST cross-correlated sample of 10149 active galaxies and analyse their positioning on the optical fundamental plane. We focus on various reasons that can cause the discrepancy between the observed FIRST radio fluxes and the theoretically expected core radio fluxes, and show that that FIRST fluxes are heavily contaminated by non-nuclear, extended components and other environmental factors. We show that the subsample of `compact sources', which should have negligible lobe contribution, statistically follow the fundamental plane when corrected for relativistic beaming, while all the other sources lie above the plane. The sample of LINERs, which should have negligible lobe and beaming contribution, also follow the fundamental plane. A combined fit of the low-luminosity AGN and the X-ray binaries, with the LINERs, results in the relation log L$_\textrm{R}$ = $0.77$ log L$_\textrm{OIII}$ + $0.69$ log $\textrm{M}$. Assuming that the original fundamental plane relation is correct, we conclude that 1.4 GHz FIRST fluxes do not trace the pure `core' jet and instantaneous nuclear activity in the AGN, and one needs to be careful while using it on the fundamental plane of black hole activity.
\end{abstract}



\section{Introduction}

The fundamental plane of black hole activity is a plane stretched out by both the stellar-mass and supermassive black holes (SMBH) in the three-dimensional space given by the mass of the black holes, the X-ray/OIII luminosities and the core radio luminosities \citep{m03,f04,s15,nb16}. The fundamental plane (FP) can be theoretically explained by the radio emission coming from synchrotron radiation produced by relativistic compact jets, and the X-ray emission originating either at a radiatively inefficient accretion flow \citep{bk79,fb95,hs03}  or at the base of the jet \citep{mar03}. This relationship suggests that the black holes of the entire mass range regulate their radiative and mechanical luminosities in the same way at any given accretion rate, when scaled to their respective Eddington rate.

The theoretical explanation of the black hole fundamental plane is based on the core component of the AGN, leaving out the extended emission (as described in \cite{f04}, based on \cite{fb95,bk79}). Hence it is required to trace the nuclear radio jet emission, to correctly constrain the fundamental plane parameters. In order to probe the `core' nuclear radio luminosity of an active galactic nuclei (AGN), it is necessary to use high-resolution and high-frequency radio observations. Namely, radio fluxes obtained with the Very Large Array (VLA) in A-configuration at 5 or 15 GHz frequencies are expected to probe the instantaneous central `core' radio flux of an AGN. However, many recent studies have used radio fluxes extracted from the Faint Images of the Radio Sky at Twenty-Centimeters \citep[FIRST,][]{w97} survey that observed with the VLA in B-configuration at 1.4 GHz \citep[e.g.][etc.]{wang06,lww08}. While some studies have obtained very different fundamental plane coefficients when non-core emission is included \citep[e.g.][etc.]{wang06}, some other studies have not found any significant deviations \citep[eg.][etc.]{wong16}.

As one moves from 15 to 1.4 GHz radio frequency, the VLA telescope beam becomes $\sim$100 times smaller. In addition to that, many of these sources have a steep spectrum ($F_{\nu} \sim \nu^{0.6}$). Hence, when the radio observation is performed at a lower frequency like 1.4 GHz, we lose a factor of $\sim$ 400 in resolving power compared to observations made at 15 GHz. Moreover, the FIRST observations are performed with the VLA in B-configuration, which has a much shorter baseline compared to the A-configuration. Considering all these factors, if we use radio fluxes reported in the FIRST survey, we lose a factor $\sim$1000 in resolving power, as opposed to the radio fluxes observed with the VLA in A-configuration at 15 GHz. Thus, it is likely that the radio fluxes reported by FIRST do not only represent the core of the AGN, since they might be contaminated by extended radio emission coming from non-nuclear sources.

Naturally, the FP studies done using the radio fluxes from the FIRST survey have reported different best-fit correlation coefficients than the original FP discovered by \cite{m03} and \cite{f04}. \cite{wang06} studied an uniform sample of all the 115 broad line AGN after cross-correlating ROSAT All-Sky Survey catalog \citep[RASS,][]{v99}, Sloan Digital Sky Survey (SDSS) survey and the FIRST catalogue. They report a much weaker dependence of the FP on the mass of the black hole than what was previously thought. They also show that the FP stretched out by the radio-loud objects have a steeper slope compared to the radio-quiet ones, whereas the previous plane relation was found to be universal for all the different types of black holes. In addition, \cite{lww08} report the same result with a much larger sample of 725 broad line AGN.

On the other hand, \cite{nb16} carefully selected only the Low-ionization nuclear emission-line regions (LINERs) in the FIRST catalogue. LINERs are expected to be the higher-mass counterparts of the low/hard state X-ray binaries (XRBs). Despite using FIRST to trace the core jet power, they find that the original FP relation holds when only LINERs are selected in the sample. This result suggests that FIRST radio fluxes can be used on the FP only in few carefully considered and specific conditions, and raises the concern of whether FIRST fluxes should be used to trace nuclear jet activities of all AGN sources.\\


For the first time, we use the FIRST radio fluxes in the `optical' fundamental plane of black hole activity. One of the main advantages of using optical emission lines as a proxy of the accretion rate is that it is measurable in a large number of sources. The aim of this paper is to check if populating the plane with FIRST fluxes is possible and if not why. The question we want to explore is whether the FIRST radio fluxes trace the core jet power and nuclear activity in the AGN, or if it is heavily contaminated by environmental and other non-nuclear factors. For this purpose, we use all the AGN in the SDSS-FIRST cross-correlated sample, with available 1.4 GHz radio luminosity, [OIII] emission line luminosity and the estimated mass of the black hole. The final sample consists of 10149 active galaxies. If this sample is indeed probing the `core' nuclear part of the AGN sources, we can heavily populate the plane and significantly refine its best-fit parameters. 

We describe the sample of 10149 active galaxies in Section 2. In Section 3, we explore the possible factors behind the discrepancy between the observed FIRST radio fluxes and the theoretically expected core radio fluxes from the FP relation. We also use crude theoretical relations to predict the extended radio luminosities of the lobes and study the positioning of different sub-samples of AGN on the optical FP.  Finally, we discuss our results and present the conclusions in Section 4.

\section{Sample selection}

The present optical fundamental plane of black hole activity uses 15 GHz radio luminosity to trace the instantaneous jet power of an AGN. Therefore it is severely limited by the low number of VLA detections of the radio core. The current FP relation is obtained with only 39 low-luminosity AGN \citep{s15}. With the aim of populating the plane and constraining its parameters we used the FIRST survey, which has the largest sample of active galaxies detected by the VLA at 1.4 GHz.\\

We selected the sample of 18286 radio-loud AGN presented in \cite{bh12} as the parent sample. They constructed this sample by combining the seventh data release of the SDSS with the NRAO VLA Sky Survey \citep[NVSS,][]{c98} and the FIRST catalogue.

As we require only those galaxies which have available [OIII] emission line fluxes, we performed a coordinate match of this parent sample with the study of emission lines in SDSS galaxies by the publicly available Gas and Absorption Line Fitting (GANDALF; Sarzi et al. 2006) code. This reduced the size of our sample to 18051 galaxies. From the whole sample, 15189 are AGN and 2853 are star-forming galaxies. We cross-correlated the 15189 AGN with the FIRST catalogue and finally selected all the active galaxies for which we have data on [OIII] emission line flux, 1.4 GHz radio flux and black hole mass (estimated from stellar velocity dispersion). This reduced the total size of the sample to 10149 AGN. The luminosity distances for these sources were derived from their redshifts reported in \cite{bh12}.

In order to include black holes of the entire mass range, we made use of the low-luminosity AGN (LLAGN) and XRB samples that were used to define the original optical fundamental plane \citep{s15}. This LLAGN sample was extracted from the Palomar Spectroscopic Survey \citep{h95}. All the 39 LLAGN with known radio luminosities at 15 GHz \citep{n05} and [OIII] line luminosities \citep{h97} were selected for the study. The stellar mass XRB sample comprises the best-studied XRBs in the hard state; GX 339-4 (\citealt{c13}), V404 Cyg (\citealt{c08}), XTE J1118+480 (\citealt{m03}) and A06200-00 (\citealt{g06}).

\section{The FIRST sample on the optical fundamental plane}

The fundamental plane of black hole activity is a universal correlation between nuclear jet power and accretion rate in black holes of the entire mass range. In order to study this plane, we selected the largest possible sample of active galaxies with available information on 1.4 GHz radio fluxes and [OIII] emission line fluxes, as well as estimates of the redshifts and the black hole masses. A sample of 10149 active galaxies is obtained by cross-correlating the SDSS catalogue and FIRST survey as described in Section 2. If FIRST 1.4 GHz radio fluxes indeed trace the instantaneous nuclear jet power, then these sources should theoretically follow the fundamental plane of black hole activity. On the contrary, a statistical disagreement between the fundamental plane relation and the observed 1.4 GHz fluxes can imply that FIRST fluxes do not purely trace nuclear jet activities, and are significantly contaminated by environmental factors and non-nuclear extended radio emissions.\\

\begin{figure*}
\centering
\includegraphics[width=172.5mm]{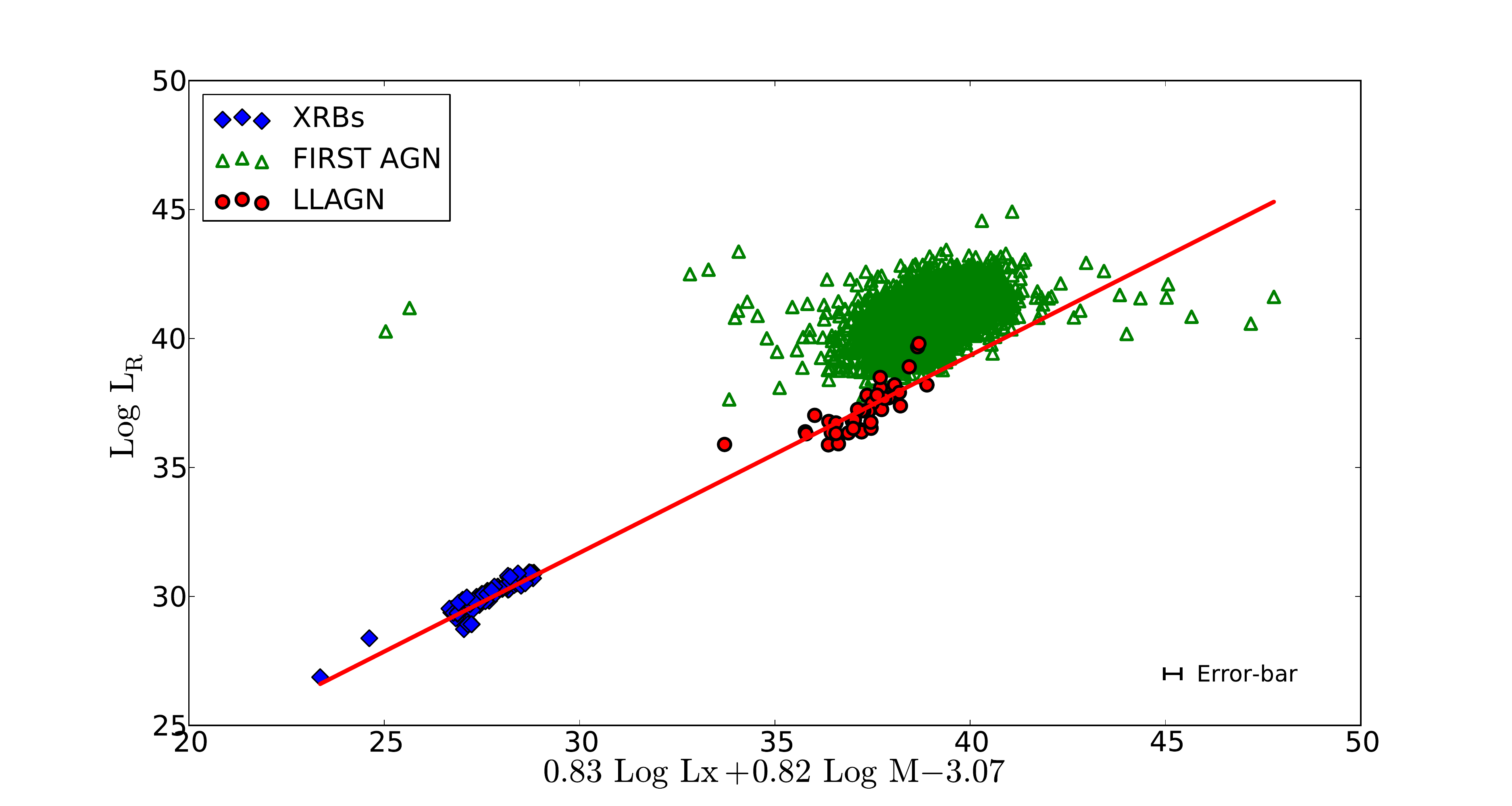}
\caption {Position of the FIRST sample on the optical fundamental plane. XRBs are shown in blue squares, LLAGN in red circles and the FIRST sources in green triangles. The red solid line is the projection of the best-fit plane using only LLAGN sample. A typical horizontal error-bar on the sources, estimated through simple error propagation using all the uncertainties involved, is shown for reference in the bottom right part of the plot. Luminosities are given in erg s$^{-1}$ while the masses are in the unit of solar mass.}
\label{fig:fp}
\end{figure*}

We use the optical fundamental plane for this study, where the nuclear [OIII] emission line is used as an indirect proxy for the accretion rate \citep{s15}. Unlike X-rays, [OIII] line luminosity is relatively easier to obtain, as it can be measured by ground-based observations. Moreover, as the [OIII] emission line originates far from the central black hole, contamination coming from relativistic beaming and the effects of torus obscuration are minimized. In addition, the [OIII] line is known to be relatively weak in metal-rich, star-forming galaxies. Hence the optical fundamental plane gives several advantages over the X-ray fundamental plane to study such objects. But it also has limitations and introduces more scatter to the fundamental plane. For example, in order to compare the active galaxies with the XRBs, a simple relation between the [OIII] and X-ray luminosities is used. This relation, log $\rm L_{3-20 keV}$/$L_{OIII}$= 2.15, has a scatter of $\sim$0.5 dex \citep{h05}, which introduces additional uncertainties in the fundamental plane coefficients.

The parameters of the original optical fundamental plane are obtained from a LLAGN sample with 15 GHz VLA A-configuration radio fluxes and a low/hard state XRB sample with 5 GHz radio fluxes \citep{s15}. In order to compare these two samples, the LLAGN 5 GHz radio fluxes were predicted from the 15 GHz VLA flux by assuming a flat radio spectrum. Here, we repeated the same procedure to estimate the 5 GHz radio fluxes for the new FIRST AGN sample observed at 1.4 GHz. Moreover, to compare the new AGN sample with the XRBs, we converted their [OIII] emission line fluxes to X-ray fluxes by using a simple linear relation, log(L$_\textrm{3-20keV}$/L$_\textrm{OIII}$) = 2.15 dex \citep{h05}.\\

In order to estimate the plane coefficients, we use the same approach of Multivariate Correlation Linear regression analysis, as described in \cite{s15}. We use the modified chi-square estimator known as merit function, which is defined as
\begin{equation}
\chi^2(a,b) = \sum\limits_i \frac{(y_i - b - \sum\limits_j a_jx_{ij})^2}{\sigma_{yi}^2 + \sum\limits_j (a_j \sigma_{x_{ij}})^2} ,
\end{equation}
where $\sigma _{x_{ij}}$ and $\sigma _{yi}$ denote the respective uncertainties, \textit{y$_i$} are the [OIII] line luminosities, \textit{x$_{1j}$} the radio luminosities at 15 GHz and \textit{x$_{2j}$} the black hole masses. The linear regression coefficients a$_j$ and the zero intercept b can be calculated by minimizing $\chi^2$.\\

We have included all the errors and uncertainties involved in each variable, in order to correctly extract the parameters of the best-fit plane with the merit function. The uncertainty in luminosity depends on both flux and distance measurements. The radio and optical fluxes typically have an uncertainty of less than 10 percent, and is fixed at 0.05 dex in our analysis. Error in the distance measurements is taken as 0.15 dex. We have used the correlation between stellar velocity dispersion and mass to estimate the AGN masses, using the M-$\sigma$ relation. Hence, the uncertainty involved in the masses is also estimated from the scatter in the M-$\sigma$ relation, which typically is around 0.34 dex \citep[eg.][]{mf01,k13}. We have incorporated all the errors mentioned here in the merit function, to correctly estimate the best-fit plane coefficients. A typical horizontal error-bar on the sources is calculated through simple error propagation using all the uncertainties mentioned above, and included for reference in the fundamental plane plots. It is important note that this is not the exact uncertainty on each of the sources, but an estimation of the expected horizontal error-bars. In addition to these, an intrinsic scatter term is included, whose exact magnitude is chosen to ensure that the reduced merit function is unity. This intrinsic scatter is generally found to be between 0.2 to 0.4 dex. It takes into account various factors like source peculiarities, non-simultaneous measurements of radio and optical flux of AGN, effects of spin, beaming statistics, absorption etc.

We note that additional uncertainty is introduced by the conversion relation between X-ray luminosity and [OIII] emission line luminosity for the AGN population \citep[$\sim$0.5 dex,][]{h05}. Although it is the main contributor in the error budget, it is not correlated with the rest of the uncertainties, and will just introduce additional scatter in the fundamental plane relation.\\

We included the new sample of FIRST active galaxies on the existing optical FP, and found that most of these sources have higher radio luminosities than predicted and lie well above the plane (see Fig. 1). The discrepancy observed between the theoretically expected and the observed FIRST radio luminosities could imply that radio-loud sources do not follow the FP of black hole activity. Alternatively, it could be explained by an over-estimation of the radio emission owing to the effects of relativistic beaming, or by the fact that the radio luminosity at 1.4 GHz has considerable contribution from non-nuclear extended radio emission which do not trace instantaneous nuclear activity. In the following subsections, we discuss each of these three possible factors in detail.

\subsection{Radio-loud sources do not follow the plane}

The first possibility to consider is that these are all powerful radio-loud sources, which do not follow the fundamental plane. This explanation is inspired by recent results from \cite{h09}, where they have studied a set of powerful radio galaxies from the 3CRR catalogue with z$<$1. They used deep XMM spectroscopy to disentangle jet and accretion-related components in the X-ray fluxes of this sample. They found that when the accretion-related X-ray component is used the sources follow the fundamental plane, whereas the sample moves above the fundamental plane when jet-related X-ray component is used.

\cite{gas11} reported similar findings by analysing a sample of 17 radio-loud LLAGN with black hole mass $ \sim10^8$M$_{\odot}$ within a narrow redshift range (0.05 $<$ z $<$ 0.11). They measured the core X-ray emission using Chandra and the radio emission with the VLA and found that almost all the sources in the sample lie above the fundamental plane. Both these papers interpreted that the sources lie above the fundamental plane whenever the X-ray emission is dominated by the jet synchrotron emission, rather than the Comptonization emission coming from the radiatively inefficient accretion flow.\\

Generally, the observed X-ray luminosity can be a superposition of several components, originating at both the accretion flow and the jet. These include inverse Compton emission from a corona, synchrotron radiation from a jet and synchrotron self-Compton emission from the accretion flow or the jet. Many of these components will change its spectral shape with the black hole mass or the luminosity. For example, the synchrotron emission from the jet may change its cut-off energy depending on the luminosity of the accretion flow \cite[see e.g.][]{ghis02}. Thus, depending on the model used for the origin of the X-ray emission, one may need to correct the spectral energy distribution of the source with respect to the black hole mass or luminosity \citep[see discussion by][]{plo12,k06}. This correction would move the sources lying above the fundamental plane back down to the correlation. We note however, that this effect is far smaller than the effect seen here for the FIRST radio sources.

Hence, although we can not completely rule out the possibility that radio-loud sources do not follow the fundamental plane, we do not expect this to significantly affect our study, especially as we are studying the optical fundamental plane, thus using [OIII] emission line luminosity in place of X-ray luminosity to trace the accretion rate of the active galaxies.

\subsection{Most of the sample is highly beamed}

\begin{figure}
\centering
\includegraphics[width=92.5mm]{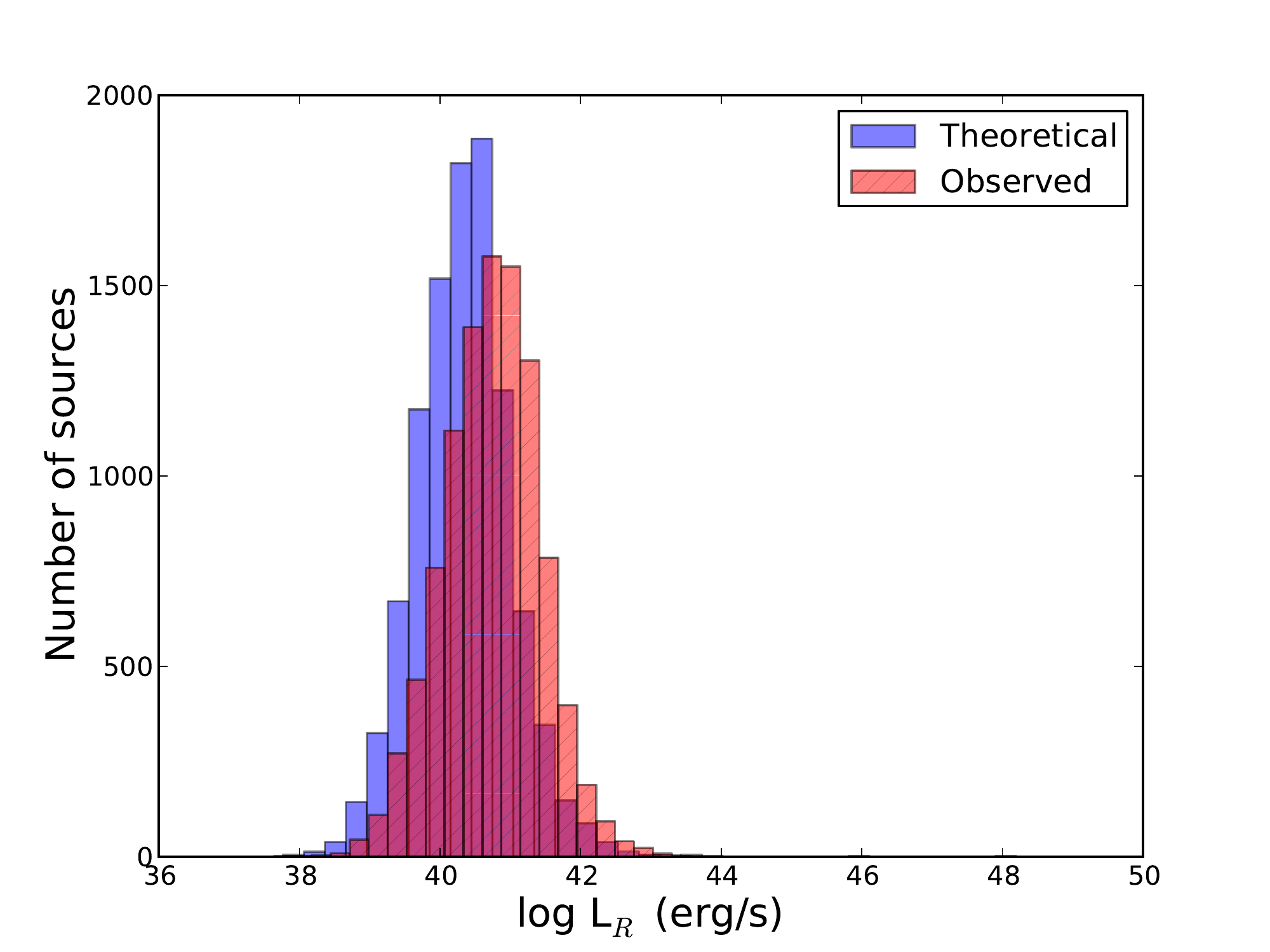}
\caption {Comparison between the radio luminosity distributions of the observed FIRST sources and the expected luminosity of the $\sim$40\% sources that could be theoretically boosted from the fundamental plane using Monte Carlo simulation.}
\label{fig:fp}
\end{figure}

The second possible physical process that can contribute towards a much higher observed radio luminosity than what is expected theoretically, is the relativistic beaming of a jet. The observed emission from relativistic jets of an active galaxy can be considerably brighter compared to the intrinsic luminosity, owing to relativistic effects.

It is not completely clear what fraction of the FIRST galaxies is relativistically beamed. \cite{lu2010} estimates the Type 1 fraction in the FIRST survey to be just 20-30$\%$. Generally, Type 2 AGNs are not beamed at all. Hence we do not expect majority of the FIRST sources to be highly boosted. In order to explore if relativistic beaming can still explain the discrepancy in the expected and the observed radio luminosity, we use a standard Monte Carlo test, as explained below.

In the first approximation, relativistic beaming can be defined by just two intrinsic parameters - the Lorentz factor and the viewing angle. If L$_\textrm{int}$ is the intrinsic luminosity and L$_\textrm{obs}$ is the observed luminosity, then $\textrm{L}_\textrm{obs} = \delta^p \textrm{L}_\textrm{int}$, where $\delta$ is the kinematic Doppler factor given as
\begin{equation*}
\delta = \frac{1}{\Gamma(1-\beta Cos\theta)}
\end{equation*}
where $\beta$ is the bulk jet velocity in units of $c$, $\Gamma$ is the Lorentz factor defined as $(1-\beta^2)^{-1/2}$ and $\theta$ is the viewing angle of the jet with respect to our line-of-sight. The exponent $p$ depends on various assumptions about jet structure, jet emission spectrum and the frequency at which the jet is observed. For a spectral index $\alpha$ defined as $S_{\nu}\propto\nu^{\alpha}$, $p$ can be approximated to be $p = 2+\alpha$ for a continuous jet or $3+\alpha$ for a jet with distinct blobs.

To check whether relativistic beaming can boost the intrinsic radio luminosity of these sources to the range of observed luminosity L$_\textrm{obs}$ , we followed the method described in \cite{s16}. We first used the optical fundamental plane relation, black hole mass and the [OIII] line luminosity, to estimate the core jet power of these sources. This is the intrinsic radio luminosity L$_\textrm{int}$ of the source. We then constructed a Monte carlo simulation to theoretically boost the intrinsic radio luminosities. This monte carlo simulation is repeated for different viewing angles and Lorentz factor distributions. We finally compared the resultant theoretical radio luminosity distribution derived from the monte carlo simulation, with the radio luminosity distribution obtained from the FIRST observations. We find that for no viewing angle or Lorentz factor distribution, the intrinsic luminosities for the whole sample can be boosted to match the observed FIRST luminosities. This rules out the possibility of relativistic beaming being the sole or the most dominant cause of the observed high radio luminosity.\\

This is generally expected because the FIRST sample contains active galaxies at various viewing angles to our line of sight. For a normal broad-line region AGN, usually $\Gamma \sim$ 5 \citep{ob82}, $\theta <$ 10$^o$ \citep{m94} and the boosting factor is expected to be less than 30, which is not enough to account for the huge discrepancy observed between the theoretically expected and the FIRST radio luminosities.

Boosting factors can go as high as 1000 only for blazars with $\Gamma <$10 \citep{ob82} and $\theta <$ 5$^o$. Blazars have a core-dominated radio morphology and their radiation is dominated by a relativistic jet oriented close to our line of sight. Hence, blazar jets can be highly relativistically beamed depending on their orientation. However, even if we assume all the 10149 sources in our sample to be highly beamed blazars and theoretically boost them from the fundamental plane, we find that only a fraction of them can be theoretically boosted to reach the luminosity range of FIRST observations. In Fig. 2 we present the comparison of the FIRST observed radio luminosity distribution with the theoretically expected radio luminosity distribution (after boosting the sources from the fundamental plane using Monte Carlo simulation). Here, we have fixed a uniform $\theta$ distribution from 0$^o$ to 30$^o$ (as higher angles are unlikely to be detected) and a power-law Lorentz factor distribution of the form $N(\Gamma) \propto \Gamma^{-2.1 \pm 0.4}$ for $\Gamma$ ranging from 1 to 40 \citep{s16}. But even if we assume that the sample is dominated by highly relativistic blazars, it is not possible to create a simulated radio luminosity distribution having statistically significant similarities with the sample observed with FIRST, for any combination of $\theta$ and $\gamma$ distributions.

Hence, although the relativistic beaming might have significant influence on the radio luminosities of these sources, we showed that this effect alone is not enough to explain the high radio luminosity observed for most of the sources. Note that in section 3.3.1, we discuss the subsample of compact sources where relativistic beaming can account for the gap between theoretically expected and observed radio luminosities. For the rest of the sample there should be other factors in addition to the effect of relativistic beaming to explain the large discrepancy between the observed and theoretically expected radio fluxes.

\subsection{1.4 GHz is dominated by extended emissions}

\begin{figure}
\centering
\includegraphics[width=92.5mm]{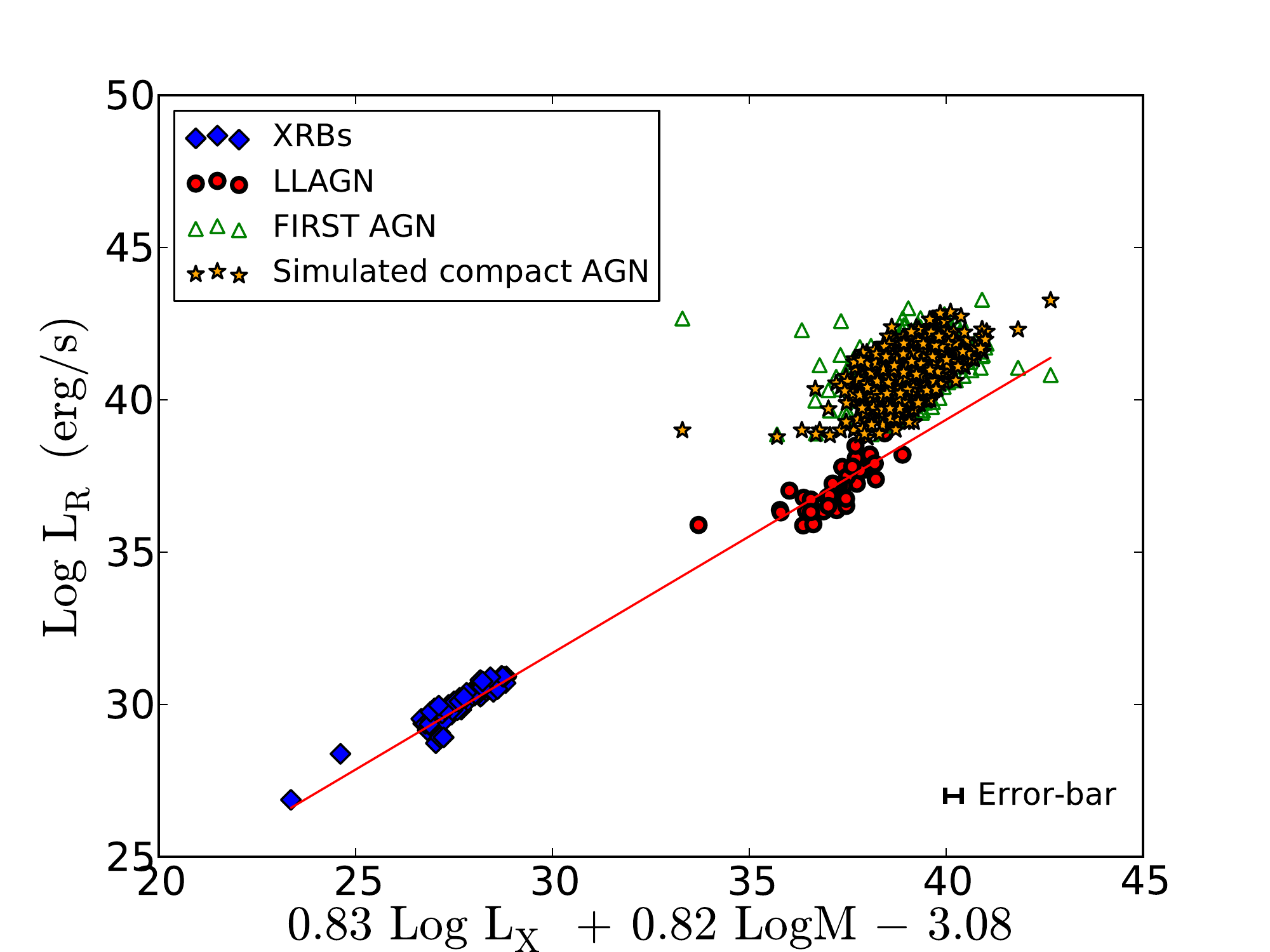}
\includegraphics[width=92.5mm]{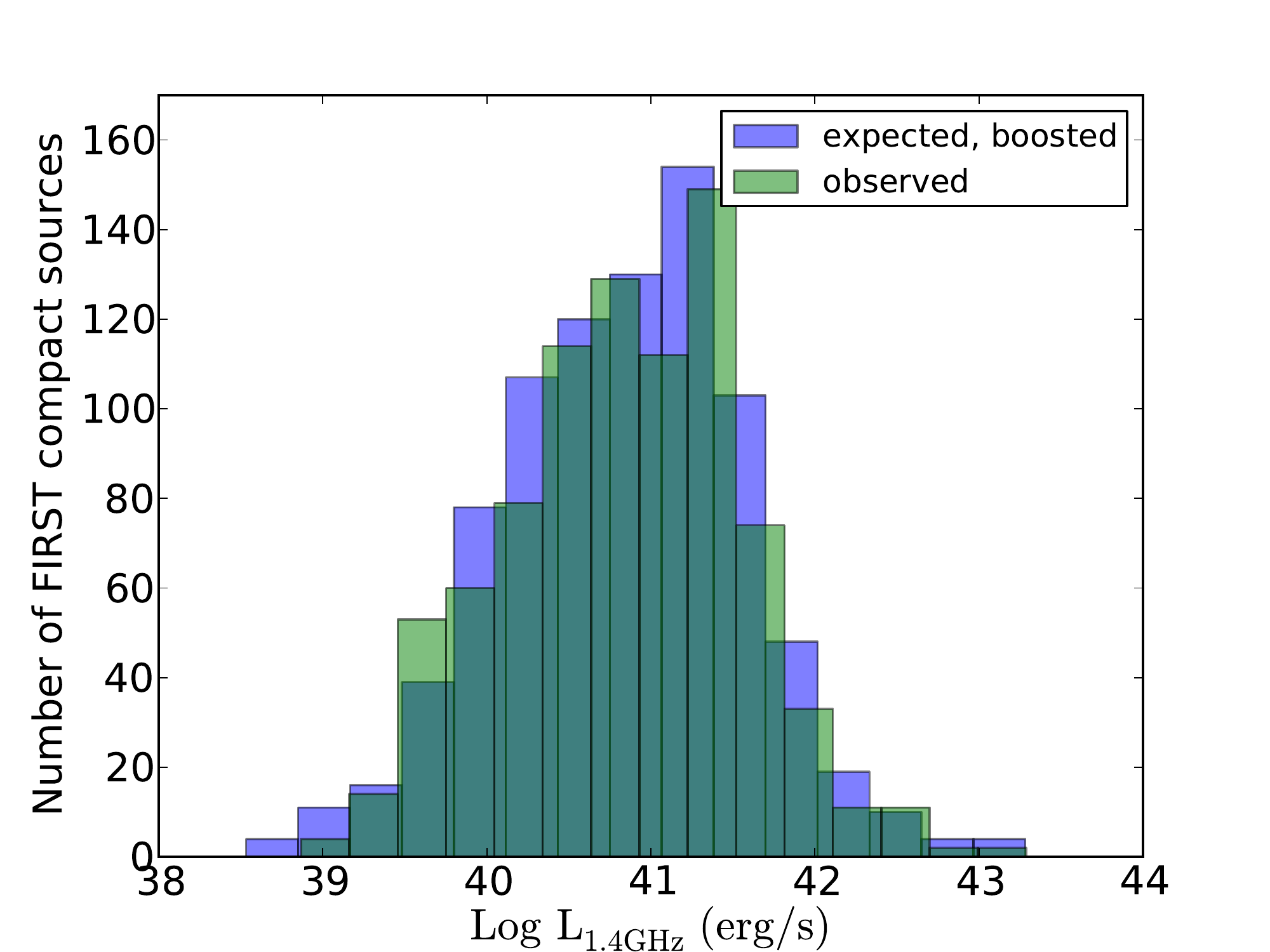}
\caption {The top panel shows the position of the 847 FIRST `compact' sample on the optical fundamental plane, as shown in green triangles. The orange squares depict the sources that are theoretically boosted from their intrinsic radio luminosity, assuming them to be dominated by a blazar population. A typical horizontal error-bar on the sources, estimated through simple error propagation using all the uncertainties involved, is shown for reference in the bottom right part of the plot. The two radio luminosity distributions are shown in the bottom panel. Luminosities are given in erg s$^{-1}$ while the masses are in the unit of solar mass.}
\label{fig:fp2}
\end{figure}

The final possibility we investigate is whether the 1.4 GHz radio flux at FIRST/NVSS resolution has considerable contributions coming from extended emissions such as the radio lobes or environmental factors that can increase the radio luminosity observed. If contribution from extended emission is indeed the dominant cause of the discrepancy between the expected core radio flux and the observed FIRST radio flux, then at least the compact sources and the LINERs should follow the optical fundamental plane.

In the following subsections we select the compact sources and the LINERs from our complete sample and study their behaviour on the optical fundamental plane.

\subsubsection{Subsample I : Compact sources}

Compact sources are the sources with negligible radio lobes and other non-nuclear extended emission. That is, if an extended radio emission contribution is the dominant reason why FIRST active galaxies are more radio luminous than expected, then we expect the compact sources to follow the fundamental plane, once corrected for the effects of relativistic beaming.

To check for this, we created a subsample including all the 847 sources for which the ratio of NVSS flux equals the flux of FIRST, after including a 5$\%$ error bar in the observed fluxes to account for the observational uncertainity in the fluxes. Both FIRST and NVSS are 1.4 GHz galaxy surveys performed with the VLA at two different configurations, resulting in different effective resolutions and sensitivities. While FIRST with its relatively high resolution underestimates the flux of extended sources and largely samples only compact sources, NVSS detects both nuclear and disk emission from galaxies. Hence, at first approximation FIRST/NVSS flux ratio can be used as an indirect measure of the compactness of the source. 

Theoretically, these `compact' sources should not have considerable lobe contribution in their FIRST fluxes and should follow the fundamental plane. But as shown in Fig 3, we find that these sources have a slightly higher radio luminosity than what is expected from the optical fundamental plane. We checked whether the relativistic beaming can explain the difference in the observed and the expected radio luminosities for these sources. For this, we followed the same approach as described in section 3.2, using relativistic boosting parameters obtained at \cite{s16}. We first use the observed black hole mass, the [OIII] line luminosity and the optical fundamental plane relation to estimate the intrinsic radio luminosity of the source. Then, we used the monte carlo simulation to introduce the relativistic beaming to the sample. In the simulation, we used a uniform $\theta$ distribution from 0$^o$ to 30$^o$ and a power-law Lorentz factor distribution of the form $N(\Gamma) \propto \Gamma^{-2.1 \pm 0.4}$ with $\Gamma$ ranging from 1 to 40 \citep{s16}, to theoretically boost the intrinsic radio fluxes of the compact subsample. In order to match the sensitivity of the simulated sample with the observed one, we incorporated the selection effect of our parent sample (radio cut-off of 5 mJy, \cite{bh12}) in the Monte Carlo simulation. As shown in the top panel of Fig 3, we observed that the `compact' subsample can be theoretically boosted to reach the exact FIRST flux range. Therefore, the compact sources in our sample that were corrected for relativistic beaming follow the fundamental plane, even when FIRST radio fluxes at 1.4 GHz frequency are used. A Kolmogorov-Smirnov (KS) test was performed to statistically quantify the similarities between the theoretically boosted and the observed radio luminosity distributions. The KS statistic is $0.05 \pm 0.01$ with a p-value of $0.18 \pm 0.13$, thereby showing that the underlying radio luminosity distributions are in agreement with being similar.\\

Thus, at least for the compact sources the 1.4 GHz radio luminosity range can be explained with a relativistically beamed AGN core. On the other hand, non-compact radio sources will have extended radio emissions, which will significantly contribute towards the 1.4 GHz fluxes measured with the VLA either with FIRST or NVSS resolution.

\subsubsection{Subsample II : LINERs}

We also studied the sub-sample of LINERs on the optical fundamental plane. LINERs are generally accepted to be powered by low-luminosity AGN, owing to the presence of broad and sometimes variable Balmer lines in their optical spectra \citep[eg.][etc.]{h97,eh01}, compact radio cores \citep[eg.][]{n05} and hard X-ray cores \citep[eg.][]{gm09}. Few recent studies also discussed the possibility of a non-nuclear origin of LINERs \citep[eg.][]{cf11,yb12}, although the results were not conclusive due to lack of full spatial or spectral information. Nevertheless, the origin of LINERs is still debatable. We are still including the LINERs in our analysis as they are known to follow the Fundamental Plane \citep[eg.][etc.]{m03,nb16}, and are seen to have a much less scatter around the plane, compared to the seyferts and the quasars \citep{k06}.

LINERs are considered to be the higher-mass counterparts of the low/hard state XRBs. They are characterized by weak and small-scale radio jets, with luminosities that are weaker than Seyfert galaxies and quasars \citep{nb16}. Accordingly, LINERs should not be considerably beamed, and are not expected to have significant lobe contributions. The sample used in the original fundamental plane have a significant fraction of LINERs, and these sources seem to have the lowest scatter around the fundamental plane \citep[see e.g.][]{k06}. Therefore, we expect the LINERs to follow the optical fundamental plane relation.

\cite{nb16} have explicitly studied LINERs on the original FP with FIRST radio fluxes. At low radio powers, the 1.4 GHz radio emission is likely to be better correlated to the core radio fluxes at 15 GHz, which are usually used in these kind of studies.  The authors report a fundamental plane parameters given by log $\textrm{L}_\textrm{R}$ = ($0.65 \pm 0.07$) log $\textrm{L}_\textrm{X}$ + ($0.69 \pm 0.1$) log $\textrm{M}$. This is significantly flatter than the original fundamental plane, reported as log L$_\textrm{R}$ = ($0.60 \pm 0.11$) log L$_\textrm{X}$ + ($0.78 \pm 0.1$) log $\textrm{M}$.

\begin{figure}
\centering
\includegraphics[width=92.7mm]{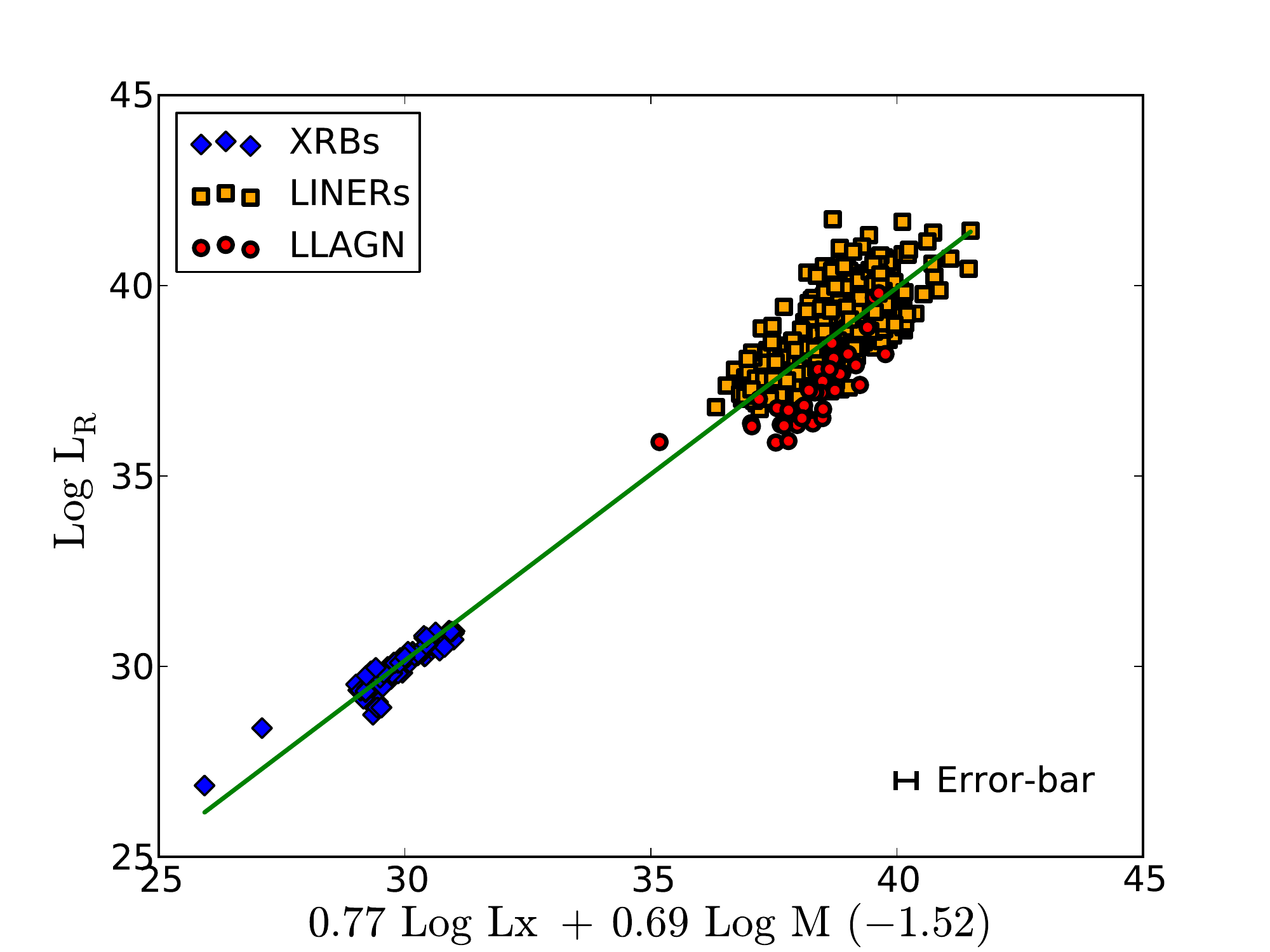}
\caption {Fitting the NISBET liners along with the LLAGN and XRBs on the optical fundamental plane. Luminosities are given in erg s$^{-1}$ while the masses are in the unit of solar mass.}
\label{fig:fp}
\end{figure}

To compare their results with the work presented here, we used the same LINER sample on the optical plane with the existing LLAGN and XRBs. As shown in Fig. 5 and 6, when plotting the LINERS using the standard fundamental plane relation, we found that the 1.4 GHz radio fluxes are consistently above the fundamental plane, and they also show a different slope. On average, brighter LINERs can deviate further from the fundamental plane, in line with a smooth transition towards the full sample of the FIRST AGN sample.  A combined fit of all the sources including the LINERs result in the relation log L$_\textrm{R}$ = ($0.77 \pm 0.04$) log L$_\textrm{OIII}$ + ($0.69 \pm 0.03$) log $\textrm{M}$ (see Fig 4). The errors in the best-fit plane coefficients are estimated using a bootstrapping routine. The fit parameters are more similar to the ones found for the optical fundamental plane, but the mass dependency is especially lower. \\

We performed statistical tests on the optical plane obtained after including the LINERs in it, and found that a simple Kendall Tau test shows that the plane correlation is significantly strong ($\tau = 0.53$, with the probability for null hypothesis as P$_{null} < 2 \times 10^{-10}$, for a sample of 847 sources, including hard-state XRBs, LLAGN and LINERs). However it is important to note that a partial Kendall Tau correlation test performed with the distance as the third variable, gives a Partial $\tau = 0.24$, with a probability for null hypothesis as P$_{null} < 7 \times 10^{-3}$, indicating that the correlation is less significant.

For comparison, when the radio luminosities are obtained with VLA in A-configuration at 15 GHz radio frequency, the significance of the plane increases. A partial Kendall Tau correlation test performed on the original fundamental plane is reported as P$_{null} < 1 \times 10^{-10}$ for the complete sample of $\sim$110 black holes \citep{m03}, while the probability of the null hypothesis for the optical fundamental plane with SMBHs is found to be P$_{null} < 3.9 \times 10^{-3}$ for 39 LLAGN sources \citep{s15}. These results show that the underlying correlation is real and not a distance-distance driven result. We can thus conclude that using 1.4 GHz radio fluxes significantly weakens the correlation between radio luminosity and the underlying core jet power.

\subsection{Various source populations on the Optical Fundamental Plane}

\begin{figure*}
\centering
\includegraphics[width=150mm]{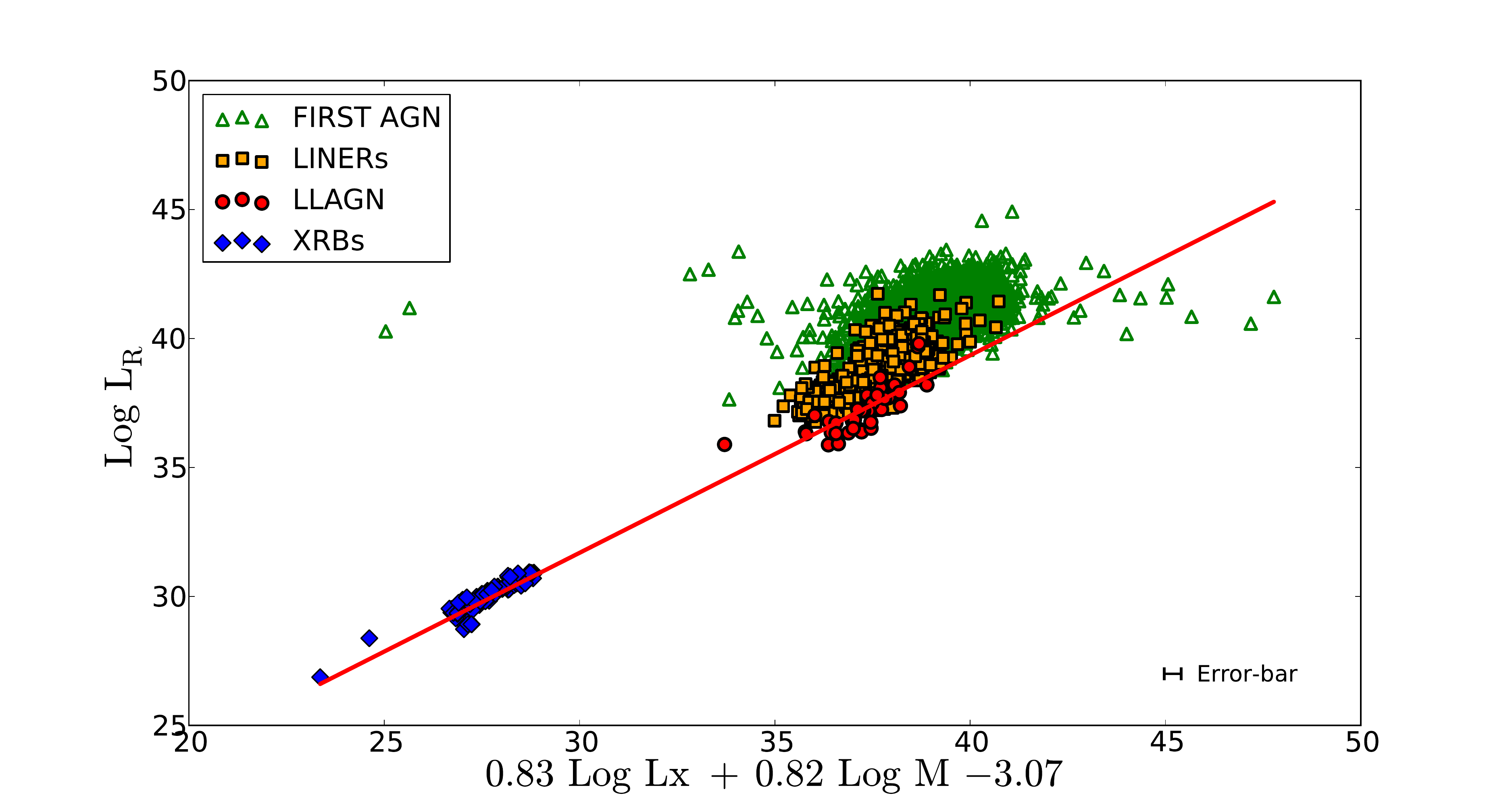}
\includegraphics[width=150mm]{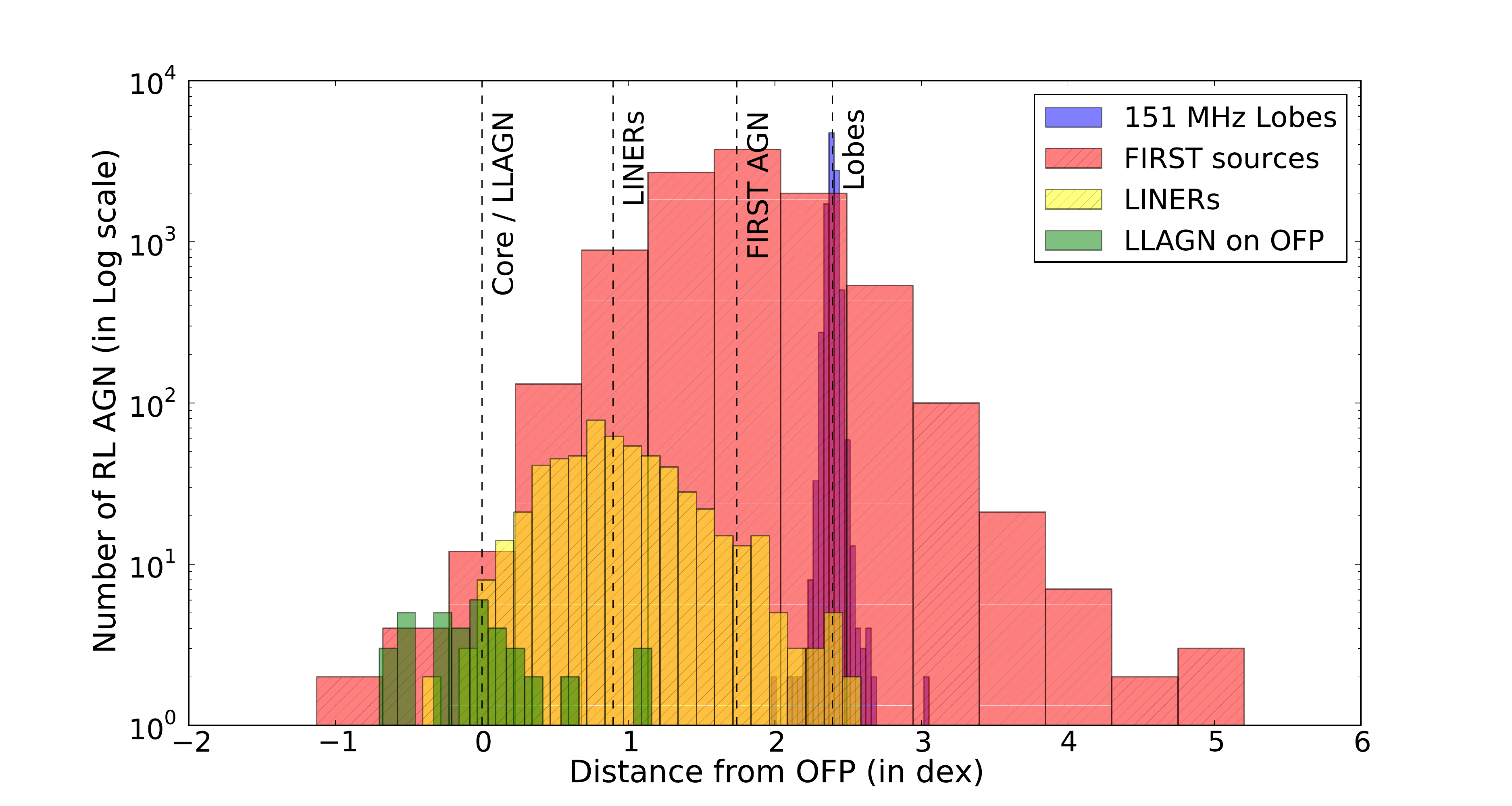}
\caption {Position and deviation of different source populations on the optical fundamental plane. In the top panel, the FIRST AGN sample (green triangles), LINERs (orange squares), LLAGN (red circles) and XRBs (blue diamonds) are plotted on the optical fundamental plane. A typical horizontal error-bar on the sources, estimated through simple error propagation using all the uncertainties involved, is shown for reference in the bottom right part of the plot. Luminosities are given in erg s$^{-1}$ while the masses are in the unit of solar mass. The bottom plot shows the distance of FIRST AGN sample (red), LINERs (yellow) and LLAGN (green) from the fundamental plane (given in dex), along with the theoretical estimation for position of the radio lobes (blue). For easy visualization of the subsamples that have comparatively fewer sources, we plot the y-axis in a logarithmic scale and show the mean position of every subsample in dashed lines.}
\end{figure*}


The core radio luminosity of an active galaxy can be estimated by the fundamental plane. The relation between the core and the lobe luminosity of an individual galaxy depends on many physical parameters like the age, size and the type of the source, as well as the physical properties of the environment. But a rough estimate of the lobe luminosity can be obtained from the jet power, using the following equation from \cite{w99}
\begin{equation*}
\textrm{P}_{\textrm{\scriptsize{jet}}} \textrm{(erg/s)} = 3 \times 10^{17} \times \textrm{f}^{3/2} \times \textrm{L}_{\textrm{\scriptsize{151MHz}}} \textrm{\footnotesize{(W Hz$^{-1}$ Sr$^{-1}$)}}
\end{equation*}
where $P_{\textrm{\scriptsize{jet}}}$ is the mechanical jet power in erg/s and $L_{\textrm{\scriptsize{151MHz}}}$ is the luminosity of the radio lobes at 151 MHz, measured in W/Hz/Sr. The factor `$f$' is a combination of all the uncertainties. For this analysis, we have fixed the value of $f$ to be $\sim 5$ \citep{d12}.

This widely used estimator of jet power from extended radio emission in active galaxies is a model-dependent predictor based on synchrotron minimum energy calculations in combination with the self-similar model of radio galaxy evolution \citep{w99}. This equation has a factor $f$ to account for the systematic errors in the model assumptions and our ignorance of true jet powers. This factor is expected to mainly depend on the type of the source and its surrounding, and is estimated to lie between 1 and 20 through observational constraints \citep{w99}. While Blundell and Rawlings (2000) finds $f$ to be $\sim$10 for a FR II population, \cite{h07} reports that $f$ lies in the range of 10-20 for FR I sources. This relation has been widely used to get an estimate the mechanical output from active galaxies based on a single low frequency luminosity measurement, with a constant value of $f$ for the entire population \cite[eg.][etc.]{h07,cb09,f011}. We use the same approach to get an rough estimate of the lobe power for our sample, with a $f$-value $\sim$ 5, following \cite{d12}.\\

We investigate different AGN populations present in our sample (LLAGN, compact AGN, LINERs and FIRST active galaxies) on the optical fundamental plane and check how they systematically deviate from the plane (see Fig 5). The optical fundamental plane is defined by the LLAGN sample, which traces the `core' radio line, while the `lobe' radio line is theoretically estimated from the equation above. The radio luminosities of the large sample of 10149 active galaxies with the FIRST observations seem to be largely distributed between the core line and the lobe line. Compared to the whole sample of FIRST active galaxies, the subsamples of compact sources and LINERs lie closer to the radio core line defined by the fundamental plane.\\

Hence we see that different AGN sources have different levels of deviations from the fundamental plane, measured as the average distance of the source population from the best-fit plane relation. The average and the range of deviation (in dex) for different source populations from the best-fit relation is shown in Fig 6, and explained below.

\begin{enumerate}
\item The LLAGN and the XRBs are used to define the fundamental plane, which estimates the `core' radio luminosity.
\item The `compact' sources present in the FIRST sample are relativistically beamed. Once the effect of beaming is removed from the radio flux, the `compact' sources statistically agree with the fundamental plane relation.
\item The LINERs lie slightly above (on average a displacement of $\sim$ 0.9 dex from the fundamental plane), when 1.4 GHz radio flux is used. But a combined fit of the LINERs with the LLAGN and the XRBs follows the trend of the fundamental plane.
\item All the remaining sources in the FIRST sample are dominated by radio fluxes coming from the lobe and other extended structures. They largely populate the area between the `core' line and the estimated `lobe' line on the fundamental plane.
\end{enumerate}

Therefore, we conclude that the 1.4 GHz radio fluxes at FIRST resolution has contribution from extended emission and lie between the radio fluxes expected from just the nuclear radio core and the extended radio lobes.

\section{DISCUSSION AND CONCLUSION}

In this paper, we used a SDSS-FIRST cross-correlated sample of 10149 active galaxies to study the optical black hole fundamental plane relationship, and investigate whether the 1.4 GHz fluxes trace the nuclear jet power. We used the 1.4 GHz radio luminosity from the VLA FIRST survey, the [OIII] emission line luminosity derived from SDSS spectra and the black hole mass obtained from stellar velocity dispersion. We showed that the active galaxies observed with the VLA at 1.4 GHz with FIRST/NVSS resolution have higher radio fluxes compared to what is expected from the fundamental plane of black hole activity or from pure nuclear core jet emission. We discussed in detail the possible reasons behind the discrepancy between the observed and theoretical radio luminosity distributions. 

We showed that the relativistic beaming alone can not explain the large difference between these two distributions, when the complete sample is considered. We investigated the subsample of 847 `compact' sources in our sample, and found that their FIRST radio luminosity range can be well explained with a relativistically beamed AGN core. Moreover, the LINER subsample, which should have negligible contribution from extended emission and relativistic beaming, also followed the fundamental plane relation. In conclusion, we may say that the FIRST fluxes of the remaining sources have considerable contributions coming from extended radio emission.

While the optical fundamental plane traces the core radio power, theoretical approximations can be used to roughly estimate the lobe radio power. We see that majority of the FIRST AGN sources lie in the flux range defined by the observed core and expected lobe radio fluxes. This is further evidence to our claim that FIRST radio fluxes have contribution from the extended radio emission, and do not purely trace the nuclear jet power. With the assumption that the fundamental plane relation is right, we suggest that one needs to be careful in using the 1.4 GHz flux obtained with FIRST/NVSS resolution to estimate the pure nuclear activity of AGN.

Furthermore, compared to the FIRST survey (with VLA in B-configuration at 1.4 GHz radio frequency), the VLA in A-configuration at 15 GHz radio observations will have a factor $\sim$1000 better in resolving power, which is necessary to remove the contribution coming from extended emission and starburst activities. If the theory behind the fundamental plane of black hole activity is right, we can use only the intrinsic `core' nuclear jet power on the fundamental plane, after removing the contributions from extended emission and relativistic beaming. Hence it is required that the radio emissions used to construct the plane are observed with high-resolution facilities (e.g. VLA in A-configuration) to resolve out all the non-nuclear extended emission and at high-frequency ($\sim$ 15 GHz) where extended emission is expected to be weaker. We conclude that it is important to be cautious when selecting the type of radio observations, and the class of AGN sources to study the fundamental plane of black hole activity and trace the nuclear jet power of a source.

\section*{ACKNOWLEDGEMENTS}
We thank the anonymous referee for insightful comments. We acknowledge Heino Falcke for the valuable discussions on the fundamental plane, and Ester Aranzana for her inputs that allowed us to improve the readability of this paper. This publication uses the NVSS and the FIRST radio surveys, carried out using the National Radio Astronomy Observatory Very Large Array. We acknowledge funding from an NWO VIDI grant under number 639.042.218.


\bsp

\label{lastpage}


\begin{thebibliography}{99}
\bibitem[Best \& Heckmann(2012)]{bh12} Best \& Heckmann 2012, MNRAS, 421, 1569
\bibitem[\protect\citeauthoryear{Blandford \& K\"{o}nigl}{1979}]{bk79} Blandford R. D. \& K\"{o}nigl A., 1979, ApJ, 232, 34
\bibitem[Brinkmann et al.(2010)]{brink00} Brinkmann W., Laurent-Muehleisen S. A., Voges W., et al. 2000, A\&A, 356, 445
\bibitem[\protect\citeauthoryear{Cattaneo \& Best}{2009}]{cb09} Cattaneo A. \& Best P. N., 2009, MNRAS, 395, 518
\bibitem[\protect\citeauthoryear{Cid Fernandes et al.}{2011}]{cf11} Cid Fernandes R., Stasinska G., Mateus A., Vale Asari N., 2011, MNRAS, 413, 1687 
\bibitem[\protect\citeauthoryear{Condon et al.}{1998}]{c98} Condon J. J., Cotton W. D., Greisen E. W., Yin Q. F., Perley R. A., Taylor G. B.  \& Broderick J. J., 1998, AJ, 115, 1693
\bibitem[\protect\citeauthoryear{Corbel et al.}{2008}]{c08} Corbel S., K\"{o}rding E. \& Kaaret P., 2008, MNRAS, 389, 1697-1702
\bibitem[\protect\citeauthoryear{Corbel et al.}{2013}]{c13} Corbel S., Coriat M., Brocksopp C., Tzioumis A.K., et al. 2013, MNRAS, 428, 2500
\bibitem[\protect\citeauthoryear{Daly et al.}{2012}]{d12} Daly R. A., Sprinkle T. B., OÕDea C. P., Kharb P. \& Baum S. A., 2012, MNRAS, 423, 2498
\bibitem[\protect\citeauthoryear{Evans et al.}{2008}]{ev08} Evans D. A., Hardcastle M. J., Lee J. C., Kraft R. P., Worrall D. M., Birkinshaw M. \& Croston J. H., 2008, ApJ, 688, 844
\bibitem[\protect\citeauthoryear{Eracleous \& Halpern}{2001}]{eh01} Eracleous M. \& Halpern J. P., 2001, ApJ, 554, 240
\bibitem[\protect\citeauthoryear{Falcke \& Biermann}{1995}]{fb95}Falcke H. \& Biermann P. L., 1995, A\&A, 293, 665
\bibitem[\protect\citeauthoryear{Falcke et al.}{2004}]{f04} Falcke H., K\"{o}rding E. \& Markoff S., 2004, A\&A, 414, 895
\bibitem[\protect\citeauthoryear{Fernandes et al.}{2011}]{f011} Fernandes C. A. C., Jarvis M. J., Rawlings S., et al., 2011, MNRAS, 411, 1909
\bibitem[\protect\citeauthoryear{Gallo et al.}{2006}]{g06} Gallo E., Fender R.P., Miller-Jones J. C. A., Merloni A. et al. 2006, MNRAS, 370, 1351
\bibitem[\protect\citeauthoryear{Gasperin et al.}{2011}]{gas11} Gasperin F., Merloni A., Sell P., Best P., Heinz S. \& Kauffmann G., 2011, MNRAS, 415, 2910
\bibitem[\protect\citeauthoryear{Ghisellini et al.}{2002}]{ghis02} Ghisellini G., Celotti A. \& Costamante L., 2002, A\&A, 386, 833
\bibitem[\protect\citeauthoryear{Gonzalez-Martin et al.}{2009}]{gm09} Gonzalez-Martin O., Masegosa J., Marquez I., Guainazzi M. \& Jimenez-Bailon E., 2009, A\&A, 506, 1107
\bibitem[\protect\citeauthoryear{Hardcastle et al.}{2007}]{h07}Hardcastle M. J., Evans, D. A. \& Croston J. H., 2007, MNRAS, 376, 1849
\bibitem[\protect\citeauthoryear{Hardcastle et al.}{2009}]{h09} Hardcastle M. J., Evans D. A. \& Croston J. H., 2009, MNRAS, 396, 1929
\bibitem[\protect\citeauthoryear{Heckman et al.}{2005}]{h05} Heckman T.M., Ptak A., Hornschemeier A. \& Kauffmann G., 2005, ApJ, 634,161
\bibitem[\protect\citeauthoryear{Heinz \& Sunyaev}{2003}]{hs03} Heinz S. \& Sunyaev R., 2003, MNRAS 343, 59
\bibitem[\protect\citeauthoryear{Ho et al.}{1995}]{h95} Ho L. C., Filippenko A. V. \& Sargent W. L. W., 1995, ApJS, 98, 477
\bibitem[\protect\citeauthoryear{Ho et al.}{1997}]{h97} Ho L. C., Filippenko A. V. \& Sargent W. L. W., 1997, ApJS, 112, 315
\bibitem[\protect\citeauthoryear{K\"{o}rding et al.}{2006}]{k06} K\"{o}rding E., Falcke H. \& Corbel S., 2006, A\&A, 456, 439
\bibitem[\protect\citeauthoryear{Kormendy et al.}{2013}]{k13} Kormendy J. \& Ho L. C., 2013, ARA\&A, 51, 511
\bibitem[\protect\citeauthoryear{Li, Wu \& Wang}{2008}]{lww08} Li Z, Wu XB \& Wang R., 2008, ApJ, 688, 826
\bibitem[\protect\citeauthoryear{Lu et al.}{2010}]{lu2010} Lu Y., Wang T., Dong X. \&Zhou H. Y., 2010, MNRAS, 404,1761
\bibitem[\protect\citeauthoryear{Maraschi \& Rovetti}{1994}]{m94} Maraschi L. \& Rovetti F. 1994, ApJ, 436, 79
\bibitem[\protect\citeauthoryear{Markoff et al.}{2003}]{mar03} Markoff S., Nowak M., Corbel S., Fender R. \& Falcke H., 2003, A\&A, 397, 645
\bibitem[\protect\citeauthoryear{Merloni et al.}{2003}]{m03} Merloni A., Heinz S. \& Matteo T. D., 2003, MNRAS, 345, 1057
\bibitem[\protect\citeauthoryear{Merritt \& Ferrarese}{2001}]{mf01} Merritt D. \& Ferrarese L., 2001, ApJ, 547, 140
\bibitem[\protect\citeauthoryear{Nagar et al.}{2005}]{n05} Nagar N. M., Falcke H. \& Wilson A. S., 2005, A\&A, 435, 521
\bibitem[Nisbet \& Best(2016)]{nb16} Nisbet \& Best 2016, MNRAS 455, 2551
\bibitem[Orr \& Browne(1982)]{ob82} Orr M. J. L. \& Browne I. W. A. 1982, MNRAS, 200, 1067
\bibitem[Plotkin et al.(2012)]{plo12} Plotkin R. M., Markoff S., Kelly B. C., Kšrding E. \& Anderson S. F., 2012, MNRAS, 419, 267
\bibitem[Saikia et al.(2015)]{s15} Saikia P., K\"{o}rding E. \& Falcke H., 2015, MNRAS, 450, 2317
\bibitem[Saikia et al.(2016)]{s16} Saikia P., K\"{o}rding E. \& Falcke H., 2016, MNRAS, 461, 2397
\bibitem[Uttley \& McHardy(2001)]{um01} Uttley P. \& McHardy I. M., 2001, MNRAS, 323, L26
\bibitem[Voges et al.(1999)]{v99} Voges W., Aschenbach B., Boller Th. et al. 1999, A\&A, 349, 389
\bibitem[Wang et al.(2006)]{wang06} Wang R., Wu XB \& Kong M., 2006, ApJ, 645, 890
\bibitem[Wong et al.(2016)]{wong16} Wong O. I., Koss M. J., Schawinski K., Kapinska A. D., Lamperti I., Oh K., Ricci C., Berney S. \& Trakhtenbrot B., 2016, MNRAS, 460, 1588
\bibitem[White et al.(1997)]{w97} White R. L., Becker R. H., Helfand D. J. \& Gregg M. D., 1997, ApJ, 475, 479
\bibitem[Willott et al.(1990)]{w99} Willott C. J,, Rawlings S., Blundell K. M. \&  Lacy M., 1999, MNRAS 309,1017
\bibitem[Yan \& Blanton(2012)]{yb12} Yan R. and Blanton M. R., 2012, ApJ, 747, 61 
\bibitem[Yuan \& Cui(2005)]{y05} Yuan F. \& Cui W., 2005, ApJ, 629, 408

\end{thebibliography}
\end{document}